\documentclass[prl,twocolumn,amsmath,amssymb,amsfonts,superscriptaddress,floatfix,showpacs]{revtex4-1}
\usepackage{graphicx}
\usepackage{subfigure}
\usepackage{dcolumn}
\usepackage{bm}
\usepackage{color}

\def\bra#1{\mathinner{\langle{#1}|}}
\def\ket#1{\mathinner{|{#1}\rangle}}
\def\braket#1{\mathinner{\langle{#1}\rangle}}

\newcommand{\etal}{\textit{et al.\/}}

\begin{document}

\title{Correlated electron behavior of metalorganic molecules: insights from density functional theory combined with many-body effects using exact diagonalization}
\author{Sumanta Bhandary}
\email{Present address: Institute of Solid State Physics, Vienna University of Technology, Wiedner Hauptstrasse 8 -10, 1040 Wien, Austria}
        \affiliation{Department of Physics and Astronomy, Uppsala University, Box 516, 751 20 Uppsala, Sweden}
\author{Malte Sch\"uler}
        \affiliation{Institute for Theoretical Physics, University of Bremen, Otto-Hahn-Allee 1, 28359
Bremen, Germany and Bremen Center for Computational Materials Science, University of Bremen, Am
Falturm 1, 28359, Bremen, Germany \\}
\author{Patrik Thunstr\"{o}m}
        \affiliation{Institute of Solid State Physics, Vienna University of Technology, Wiedner Hauptstrasse 8 -10, 1040 Wien, Austria}
\author{Igor di Marco}
        \affiliation{Department of Physics and Astronomy, Uppsala University, Box 516, 751 20 Uppsala, Sweden}
\author{Barbara Brena}
        \affiliation{Department of Physics and Astronomy, Uppsala University, Box 516, 751 20 Uppsala, Sweden}
\author{Olle Eriksson}
         \affiliation{Department of Physics and Astronomy, Uppsala University, Box 516, 751 20 Uppsala, Sweden}
         \author{Tim Wehling}
        \affiliation{Institute for Theoretical Physics, University of Bremen, Otto-Hahn-Allee 1, 28359
Bremen, Germany and Bremen Center for Computational Materials Science, University of Bremen, Am
Falturm 1, 28359, Bremen, Germany \\}

\author{Biplab Sanyal}
\email{Biplab.Sanyal@physics.uu.se}
	\affiliation{Department of Physics and Astronomy, Uppsala University, Box 516, 751 20 Uppsala, Sweden}

\begin{abstract}
A proper theoretical description of electronic structure of the 3d orbitals in the metal centers of functional metalorganics is a challenging problem. In this letter, we apply density functional theory and an exact diagonalization method in a many body approach to study the ground state electronic configuration of an iron porphyrin (FeP) molecule. Our study reveals that  dynamical correlation effects are important, and FeP is a potential candidate for realizing a spin crossover due to a subtle balance of crystal field effects, on-site Coulomb repulsion and hybridization between the Fe d-orbitals and ligand N p-states. The mechanism of switching between two close lying electronic configurations of Fe-d orbitals is shown. We discuss the generality of the suggested approach and the possibility  to properly describe the electronic structure and related low energy physics of the whole class of correlated metal centered organometallic molecules.
\end{abstract}

\maketitle

Molecular magnets combine low dimensionality and inherent confinement effects with strong electron correlations and hold prospects in the context of spintronics. An important molecular property is bistability, i.e., the possibility of realizing two different spin states, which can in principle be accessed and manipulated externally. This is important as the switching of spin has a pronounced effect on measurable quantities, like magnetic anisotropy and spin dipole moment contribution \cite{sumantaprl}. Finding ways for the efficient manipulation of the magnetic state \cite{spinswitch1,sumantachemi,sumantaprl, nsr} of transition metal (TM) centered porphyrin (TM-P) and phthalocyanine (TM-Pc) molecules have critical consequences in this regard. A crucial interplay between molecular ligand field and spin pairing energy makes only a subspace of this class of materials to respond to spin crossover.

The magnetic properties in TM-P/TM-Pc are largely governed by the metal center, which features  sizable local Coulomb interactions ($U\sim 4$~eV and $J\sim 1$\,eV) and is simultaneously subjected to crystal fields, spin-orbit coupling and orbitally dependent hybridization with the ligands. Electronic correlations are expected to arise \cite{Fazekas} and the description by local density approximation (LDA) or generalized gradient approximation (GGA) thus potentially becomes inadequate, for example, leading to underestimated or even vanishing HOMO-LUMO (HOMO=Highest Occupied Molecular Orbital and LUMO=Lowest Unoccupied Molecular Orbital) gap. Hence, the treatment of the molecular electronic structure in terms of correlated electron theories like ligand field theories or Anderson impurity models become crucial. These model based approaches can be very helpful to trace the physical origin of phenomena like spin-state switching \cite {Fazekas}, emergence of magnetic anisotropies \cite{bi2se3} or many body resonances \cite{hui} as soon as solid links between model and the realistic structure can be established.


In this letter, we have adapted a hybrid approach \cite{dft++} (DFT++), which links Density Functional Theory (DFT) and Anderson's impurity model to study physical properties of FeP and FePc. We demonstrate how the interplay of Coulomb interactions, crystal fields and hybridization with the ligands, which are fully captured in our theory, lead to correlated electron physics, and how this theory describes the  $S=1 \to S=2$ spin-crossover in the Fe$^{2+}$ metal center of FeP. Furthermore, the crossover between different close lying ground state electronic configurations within the $S=1$ subspace of the correlated Fe-d orbitals is analyzed.

The problem can be cast in the following way. The delocalized orbitals in the organic ring in FeP are described by LDA or GGA but the Fe center is considered as an impurity embedded in the organic host and is described by Anderson's impurity model \cite{anderson}. 
%

\par The model Hamiltonian  for the impurity problem can be expressed as :
\begin{eqnarray}
             H=\sum_{i,j}\epsilon^{d}_{ij}d^{\dagger}_id_j+\frac{1}{2}\sum_{i,j,k,l}U_{ijlk}d^{\dagger}_id^{\dagger}_jd_kd_l \nonumber \\
          +\sum_{ik}(V_{ik}c^{\dagger}_kd_i+H.c..)+\sum_{k}\epsilon_kc^{\dagger}_kc_k, \label{eq:ham}
\end{eqnarray}
where $\epsilon^{d}_{ij}$ describes the onsite energies and ${i,j,k,l}=(m,\sigma)$ represents combined orbital and spin indices. $d_i$ is the annihilation operator while $U_{ijlk}$ represents the local Coulomb interaction. $U_{ijlk}$ is parametrized by Slater parameters. We have chosen U=4 eV and J=1 eV for $l=2$ (3d-orbitals) in our calculations. In the above equation, the first two terms represent electrons in Fe 3d orbitals. The third term describes the interaction with the surrounding atoms while the fourth term is for the delocalized ligand electrons with energies $\epsilon_k$. The hopping matrix element $V_{ik}$ appears in the hybridization function, which is represented as :
\begin{equation}
             \Delta_{ij}(\epsilon)=\sum_{k}\frac{V_{ik}V_{kj}}{\epsilon+i\delta-\epsilon_k} \label{eq:hyb}.
\end{equation}
The energy dependent hybridization function can be obtained from first principles calculations. We follow the approach considered by Karolak \etal \cite{dft++} to construct the local Green function from DFT. The Kohn-Sham Green function $G_{KS}$ can be calculated from the Lehmann representation \cite{economou} using
 \begin{equation}
             G_{KS}(\epsilon)=\sum_{nk}\frac{\ket{\psi_{nk}}\bra{\psi_{nk}}}{\epsilon+i\delta-\epsilon_{nk}},
\end{equation}
where $\psi_{nk}$'s and $\epsilon_{nk}$'s are the Kohn-Sham eigenstates and eigenvalues for band $n$ and reciprocal space point $k$. The projection of the full Green's function to an atom centred local Green's function $G^{mm^{'}}_{imp}$ for localized orbitals is needed, which in our case are cubic harmonics ($\chi_{m}$). 
\begin{equation}
             G^{mm^{'}}_{imp}(\epsilon)=\sum_{nk}\frac{P^{m}_{nk}{P^{m^{'}}_{nk}}^{*}}{\epsilon+i\delta-\epsilon_{nk}}
\end{equation}
 where $P^{m}_{nk}=\braket{\chi_{m} \mid \psi_{nk}}$ and ${P^{m^{'}}_{nk}}^{*}=\braket{\psi_{nk}\mid \chi_{m'}}$. The hybridization function is calculated from the local impurity Green's function from the expression :
 \begin{equation}
            G^{-1}_{imp}(\epsilon)=\epsilon+i\delta-V_{cryst}-\Delta(\epsilon)
            =\epsilon+i\delta-\tilde{\Delta}(\epsilon).
\end{equation}
In the above expression, $G_{imp}$ is the projected Green's function on  local orbitals and $\tilde{\Delta}$ combines the hybridization function $\Delta$ and  static crystal field $V_{cryst}$ . If the bath orbitals (defined by $\epsilon_k$ and $V_{ik}$ in Eqn. (\ref{eq:hyb})) are limited to a small number of discrete orbitals only, the many body problem defined in Eqn. (\ref{eq:ham})  can be solved by means of exact diagonalization (ED), which will be used here.

\begin{figure}[h]
\begin{center}
\includegraphics[width=9cm]{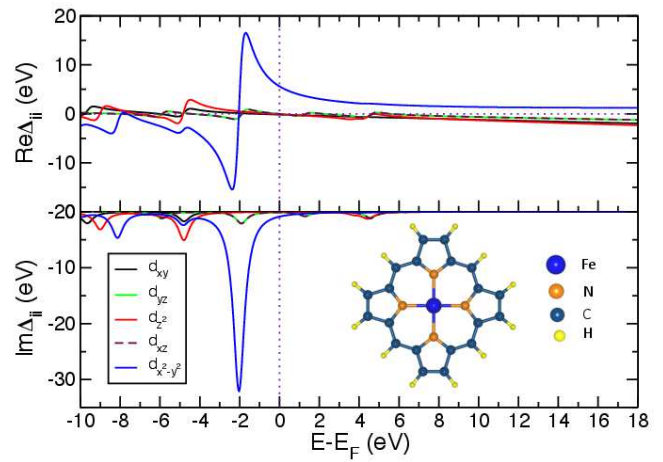}
\end{center}
\caption{\label{fig1} (Color online) Real and imaginary parts of the hybridization function for Fe in FeP
calculated with PBE in the non-spin polarized mode. The geometry of FeP is shown in the inset with the atoms labeled by their
types.}
\end{figure}

Electrons in Fe-3d orbitals hybridize mostly with the orbitals of the four surrounding N-atoms, that provide a square planar ligand field. The effect of the outer C-ring is rather indirect as that mainly shifts N-p levels by both in-plane and $\pi-\pi$ interaction. The energy dependence of that interaction with surrounding N-atoms can be described by an energy-dependent hybridization function $\tilde{\Delta}(\epsilon)$. We have employed non-spin polarized density functional calculations within local density and generalized gradient approximations to extract the hybridization functions. The DFT calculations were performed using the VASP code \cite{vasp} that employs a plane wave basis and projector augmented wave method. In Fig.~\ref{fig1}, real and imaginary parts of $\tilde{\Delta}$ are shown. The imaginary part of Im$\tilde{\Delta}$ quantifies the density of bath states coupling to each impurity orbital weighted by the hybridization matrix elements V$_{ik}$.  As seen from Fig.~\ref{fig1}, the most dominant peak in Im$\tilde{\Delta}(\epsilon)$  is observed for the Fe-d$_{x^2-y^2}$ orbital at 2.04 eV below the Fermi energy. The formation of in-plane $\sigma$ bonds with axial N-ligands explains this pronounced peak. It should be noted that the other in-plane orbital d$_{xy}$ shows almost no hybridization apart from a small peak at 4.8 eV below the Fermi energy. The out-of-plane orbitals have relatively small peaks in Im$\tilde{\Delta}$, among which the one closest to the Fermi energy is of d$_\pi$ character at -2 eV. Appearance of this peak reflects a $\pi-\pi$ interaction of Fe d$_{\pi}$ (d$_{xz}$, d$_{yz}$) orbitals with N-p$_z$ orbitals, which is expected in the square planar ligand field of the FeP molecule. The other out-of-plane and in-plane contributions are present in the -4.5 to -10 eV energy range. Taken together, the hybridization function reveals a strong in-plane interaction between Fe-d$_{x^2-y^2}$ orbitals and the N-p ligand states along with a much weaker interaction amongst all other orbitals.

The real part of the hybridization function Re$\tilde{\Delta}$ describes the energy dependent ligand field,  combined with a static crystal field, which can be obtained at the $\epsilon \to \infty$ limit. The strong resonance with the host and d$_{x^2-y^2}$ orbital manifests itself also around -2.04 eV in Re$\tilde{\Delta}$. This strong ligand field pushes d$_{x^2-y^2}$ high in energy causing almost no occupancy in either spin channel. In the gas phase, six electrons in Fe$^{2+}$, thus, are distributed in the remaining four d-orbitals, with four and two electrons in majority and minority spin channels, respectively, giving rise to an intermediate spin state (S=1).  

An intriguing phenomenon of strain induced spin state switching is observed theoretically in FeP \cite{sumantachemi, sumantaprl, liao}, which is relatively difficult to obtain for other molecules with other TM centers \cite{kahn, gutlich} or different structures, e.g., transition metal phthalocyanines. Iron Phthalocyanine (FePc), for example, has an Fe center but along with a larger organic ring. This results in a stronger hybridization of Fe with the neighboring N atoms, which is reflected in the appearance of a d$_{x^2-y^2}$ -peak in Im$\tilde{\Delta}$ at -1.85 eV and a stronger hybridization, as shown in the Supplementary Information (SI). The shift of the bath energy to -1.85 eV for FePc compared to -2.04 eV in FeP is due to the presence of four extra N atoms connected to the pyrolle rings in FePc (See Fig. 2 in SI). These four N atoms are not directly bonded to Fe but are connected to N atoms in the Fe-N$_4$ block via C atoms.

To quantify the conditions required for spin state switching of FeP, we have varied crystal field and hybridization strength within DFT++ method. The variation of these parameters mimics the strain effect on the molecule. Also as mentioned in the previous section we found that the crystal field splitting of d$_{x^2-y^2}$ is larger compared to other orbitals and as will be discussed below, this splitting is responsible for the spin switching. For this part of the calculation we have kept a common reference level of the rest of the orbitals separated from d$_{x^2-y^2}$ by V$_{cryst}$, which was varied. An independent variation of the hopping matrix elements, $V_{d_{x^2-y^2}}$, is done along with the  variation of V$_{cryst}$. One needs to keep in mind that for our model calculation,  V$_{cryst}$ and $V_{d_{x^2-y^2}}$ are independent, while in DFT calculation (discussed in the latter part of this section), these parameters are implicitly related. The ligand field variation due to the strain effect simultaneously changes V$_{cryst}$ and $V_{d_{x^2-y^2}}$.  For the remaining part of the discussion we will refer to $V_{d_{x^2-y^2}}$ as V, as this the only bath-site coupling considered for our model calculations. 

\begin{figure}[h]
\begin{center}
\includegraphics[width=9cm]{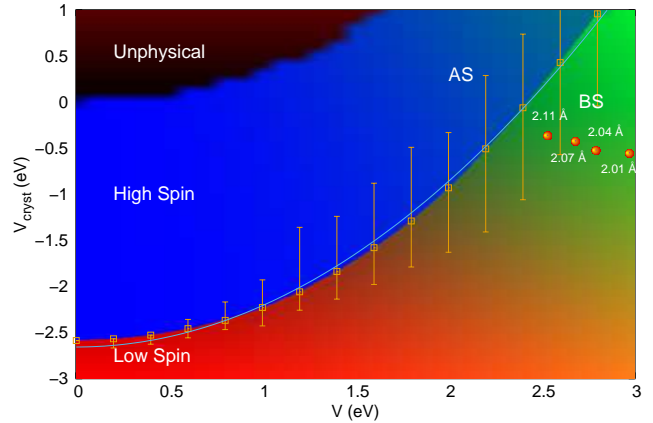}
\end{center}
\caption{\label{fig2} (Color online) The phase diagram depicting the spin states of FeP with the tuning of the static crystal field V$_{cryst}$ and the hybridization strength V. The phase boundary is accompanied by the allowed values of V$_{cryst}$  for fixed values of V. The blue curve is a result of fitting with a tight-binding model described in the Supplementary Information. AS and BS indicate antibonding and bonding regions. Calculated values of V$_{cryst}$ and V from DFT (non spin-polarised PBE) are shown in orange circles along with the corresponding Fe-N bond lengths.}
\end{figure} 

Fig.~\ref{fig2} depicts the spin phase diagram of the central atom in FeP. The phases are defined by the characteristic energy contributions and demonstrated  with RGB color code (see supplementary information). For V $\to$ 0 and with sufficiently high crystal field (V$_{cryst}$ $\textgreater$ 2.6 eV), 6 electrons occupy the degenerate d$_{rest}$ level. Hence, the ground state will be a low spin state with an energy gain from crystal field (red) but with the cost of exchange energy (blue). The ground state attains a high spin state for low crystal fields. A strong V pushes the molecular orbital containing predominantly d$_{x^2-y^2}$ and hence, required crystal field for spin crossover is small. The intermediate spin state, in this situation is governed by the Fe-N bonding (green). Crystal field can be tuned to occupy anti-bonding molecular orbital and hence, S=2 spin state can be achieved (AS). The dependence of V with V$_{cryst}$ is  follows a quadratic behaviour. The fitted curve in Fig.~\ref{fig2} at the phase boundary is obtained from a mean field model (see supplementary information) containing one particle Hamiltonian but with a renormalized onsite energy $\epsilon^d$. The discrepancy between the phase boundary and the fitted curve lies in proper renormalization of  $\epsilon^d$, that depends linearly with $N$, by electron correlation energy, which has a dependence of $N^2$.  

As the spectrum of the FeP molecule is gapped, we placed chemical potential in the middle of the gap. But   it is also possible to vary the average onsite d-energy $\epsilon_d$, while keeping the total number of particles in the system fixed. 
In addition, the average $d$-electron on-site energy is not exactly known from the DFT simulations --- a problem usually referred to as "double counting problem" in DFT++, DFT+U, or DFT+DMFT approaches. In our case, this means that the average on-site energy or more precisely the average energy difference between the bath level and the Fe $d$-block carries some uncertainty. We thus vary $\epsilon_d$ in a range, while keeping the total number of particles in our system constant. In this way, the high-spin to low-spin transition line has error bars associated with some of the calculated points shown in Fig.~\ref{fig2}.

At a stronger coupling V,  the shifting of the onsite energy $\epsilon_d$ will allow a variation of V$_{cryst}$, where a spin crossover (SCO) can happen. The range of V$_{cryst}$ due to the variation of $\epsilon_d$ is presented by the red error bars in Fig.~\ref{fig2}. In the  regime of weak hybridization, the system is described predominantly by crystal fields. There are in particular, no charge fluctuations to $N=7$ or $N=5$ Fe impurity states, which are generally affected by changing $\epsilon_d$. As there is no mixing with $N\neq 6$ impurity states in the limit of $V\to 0$, the error bar is vanishingly small in this limit. For V beyond 2.8 eV, this particular scenario of crystal field will not be able to switch the spin state. An orbital reversal, i.e., the d$_{ x^2-y^2}$ energy becoming lowered compared to other orbitals is needed in this case, which requires a different kind of charge distribution in the molecule. Thus a transition between S=1 and S=2 states could be realized for V $<$ 2.8 eV while S=1 should be obtained generally for V $>$ 2.8 eV. 

\begin{figure}[h]
\begin{center}
\includegraphics[scale =0.32]{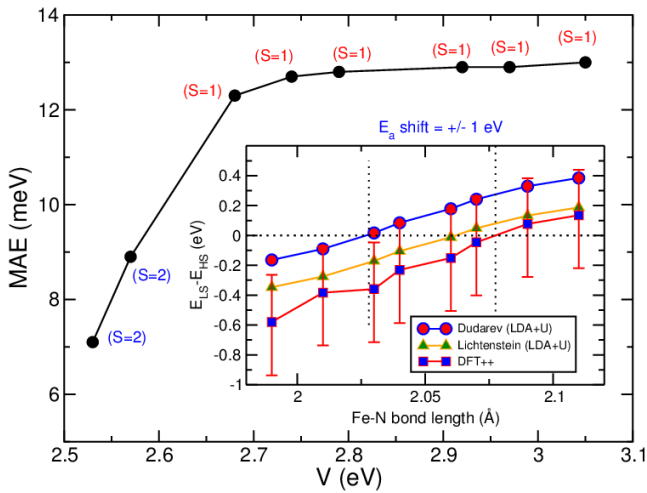}
\end{center}
\caption{\label{fig3} (Color online) Magnetic anisotropy energy (MAE) as a function of hybridisation V. The corresponding spin states of the ground state configurations are indicated. (inset) Energy differences between low spin (LS) and high spin (HS) states calculated by DFT++ and PBE+U (double counting by Dudarev and Lichtenstein) methods as a function of Fe-N bond lengths in FeP are shown. The error bars correspond to the variation of onsite energy (model Hamiltonian in eq. 1) in DFT++ calculations. }
\end{figure} 

Weakening of the ligand field leading to a spin state change, however, needs a nontrivial chemical or physical procedure. The existence of high spin (S=2) porphyrin complexes with $d^6$ configuration has  so far been observed in non-planar molecules with five or six coordination of the central Fe atom\cite{s2}. Higher coordination leads to out of plane shift of the central Fe atom (five coordination) or symmetrical Fe-N block expansion, resulting in a weaker ligand field. For four coordination, however, the S=2 state is yet to be observed experimentally in gas phase or on a surface. Fig.~\ref{fig2} establishes the parameter space for when this state is to be expected.   

The variation of V is also studied with PBE by varying the Fe-N bond length in the molecule. The highest value of V (3.16 eV) is obtained for a Fe-N bond length of 1.97 \AA. The four data points, shown in Fig.~\ref{fig2} in filled orange circles, are for bond lengths of 2.00, 2.04, 2.07 and 2.11 \AA, respectively.  As mentioned in Refs.~\onlinecite{sumantaprl, sumantachemi} for FeP either physisorbed or chemisorbed on surfaces, the required bond length of Fe-N in FeP for spin switching is beyond 2.03 \AA~ within PBE+U approximation with U$_{eff}$=3 eV. \cite{marzari} This is in agreement with the data shown in Fig.~\ref{fig2}, where both DFT++ and PBE suggest a spin crossover beyond 2.04 \AA. A detailed comparison between the bond lengths required for spin-crossover in PBE+U and DFT++ is shown in Fig.~\ref{fig3}. A comparison of PBE and LDA calculations yield values of the static crystal field, bath energy and hybridisation as 0.62(0.65) eV, -2.04(-1.9) eV and 3.16(3.39) eV for PBE(LDA). These values indicate that the two approximations for exchange-correlation functional do not show pronounced difference.

In Fig.~\ref{fig3}, calculated magnetic anisotropy energies (MAE) within DFT++ are shown for different strengths of V. The MAE is defined as the energy difference between the lowest many-body eigenstate with out-of-plane magnetic moment and the ground state which has in-plane oriented moment. It is observed that for low values of V (large Fe-N bond lengths), the MAE decreases with decreasing V whereas for large V, a more or less constant value of MAE is obtained. In all the cases, we have found the easy axis of magnetization to lie in the plane of the molecule. In the inset of Fig. 3, the spin-crossover properties of FeP as predicted by different flavors of PBE+U and DFT++ is compared. In the PBE+U calculations, we stabilize the non-favorable spin solutions by constraining the spin moments. It is seen from the figure that there is a notable difference between two PBE+U methods (Dudarev and Lichtenstein) in the Fe-N bond length required for LS-HS transition, while the results obtained from Lichtenstein PBE+U and DFT++ methods are close to each other. This is understandable since both, the Lichtenstein variant of PBE+U and DFT++, employ the full four fermion Coulomb matrix as defined through the Slater parameters, whereas Dudarev assumes a simplified Coulomb vertex.

\begin{figure}[h]
\begin{center}
\includegraphics[width=9cm]{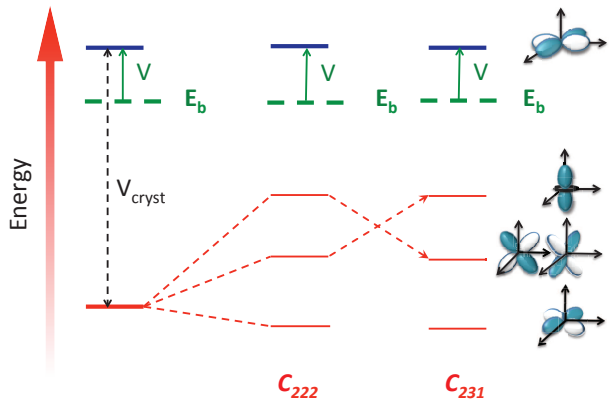}
\end{center}
\caption{\label{fig4} (Color online) Schematic diagram depicting the C$_{222}$ and C$_{231}$ configurations.
V$_{cryst}$ describes the energy separation between d$_{x^2-y^2}$ and the rest of the d-orbitals treated as degenerate. V denotes the strength of hybridization between the d$_{x^2-y^2}$ and the bath state with an energy E$_{b}$. The specific orbitals corresponding to the occupation of the energy levels are also indicated.}
\end{figure} 

The free molecule spin state (S=1) is particularly interesting from the point of view of ground state configuration. A strong ligand field in the free molecule leaves d$_{x^2-y^2}$ nearly unoccupied and  an intermediate spin state (S=1) can have ground state electronic configuration with  2 electrons in d$_{xy}$, 3 electrons in d$_{\pi}$ and 1 electron in d$_{z^2}$. We will refer to  this configuration a $C_{231}$ (d$_{xy}^{2}$d$_{\pi}^{3}$d$_{z^2}^{1}$) configuration. The other possible configurations within the S=1 ground state multiplet are $C_{222}$ (d$_{xy}^{2}$d$_{\pi}^{2}$d$_{z^2}^{2}$), $C_{141}$ (d$_{xy}^{1}$d$_{\pi}^{4}$d$_{z^2}^{1}$) and $C_{132}$ (d$_{xy}^{1}$d$_{\pi}^{3}$d$_{z^2}^{2}$) among which the $C_{222}$ appears to be very close in energy to the energy of  $C_{231}$. From Re$\Delta$, a splitting can also be seen among $d_\pi$, d$_{z^2}$ and d$_{xy}$ orbitals but in a relatively small energy scale.  To acquire a clear view of the ground state configuration, we varied the crystal field in presence and absence of coupling V. At the first step we only considered a crystal field splitting  between d$_{x^2-y^2}$ and the averaged position of remaining orbitals, as shown in the left most part of Fig.~\ref{fig4}. In this pure crystal field situation, the S=1 state leads to the $C_{231}$ ground state configuration. As the coupling V is switched on, the ground state configuration becomes onsite energy dependent. This happens because the inclusion of strong hybridization will admix N=5 and N=7 impurity occupancies with the pure N=6 ground state configuration of the pure crystal field situation. Varying on site energy, the ground state configuration can be modified. The energy scale associated to this change is of the order of meV. A relative splitting among the remaining orbitals has even more pronounced effects in determining the ground state configuration. As shown in Fig.~\ref{fig4}, with additional splitting, if d$_{\pi}$ stays above d$_{z^2}$, $C_{222}$ is stabilized. However, in the reversed situation, $C_{231}$ is obtained. This change occurs in at least one or two order higher energy scale compared to the many-body effects induced configuration change, revealing the crystal field to be the dominant factor.

In summary, we have presented ground state electronic properties of Fe in FeP molecule with a hybrid approach of DFT combined with a many body treatment, using exact diagonalization. We have demonstrated that a delicate interplay of the static crystal field, ligand hybridization and Coulomb interactions promotes iron porphyrin to be a potential candidate for realizing spin-crossover behavior. In general, our calculated phase diagram indicates the possibility of tuning electronic and magnetic properties of organometallic molecules to serve the purpose of molecular electronics and storage devices. Moreover, the long-standing debate regarding the electronic ground state configuration, e.g., $C_{222}$ vs. $C_{231}$ has been solved by identifying the proper parameters required to switch one to the other, which will have important consequences for the spin dipole moments and magnetic anisotropies, where the energy positions of d-orbitals with specific symmetries are important.

B. S. acknowledges Carl Tryggers Stiftelse for financial support. O.E. Acknowledges support from VR, the KAW foundation, eSSENCE. We acknowledge supercomputing allocation from Swedish National Infrastructure for Computing. T.O.W. and M. S. acknowledge support from the Central Research Development Fund of the University of Bremen and the DFG via FOR 1346.

\end{document}